\documentclass[8.5pt,twoside,twocolumn]{article}
\usepackage{times,mathptmx}
\usepackage{sectsty}
\usepackage{balance} 

\usepackage{graphicx} 
\usepackage{lastpage}
\usepackage[format=plain,justification=raggedright,singlelinecheck=false,font=small,labelfont=bf,labelsep=space]{caption} 
\usepackage{fancyhdr}

\begin{document}

\renewcommand{\thefootnote}{\fnsymbol{footnote}}
\renewcommand\footnoterule{\vspace*{1pt}%
\hrule width 3.4in height 0.4pt \vspace*{5pt}} 
\setcounter{secnumdepth}{5}

\makeatletter 
\def\subsubsection{\@startsection{subsubsection}{3}{10pt}{-1.25ex plus -1ex minus -.1ex}{0ex plus 0ex}{\normalsize\bf}} 
\def\paragraph{\@startsection{paragraph}{4}{10pt}{-1.25ex plus -1ex minus -.1ex}{0ex plus 0ex}{\normalsize\textit}} 
\renewcommand\@biblabel[1]{#1}            
\renewcommand\@makefntext[1]%
{\noindent\makebox[0pt][r]{\@thefnmark\,}#1}
\makeatother 
\renewcommand{\figurename}{\small{Fig.}~}
\sectionfont{\large}
\subsectionfont{\normalsize} 

\renewcommand{\headrulewidth}{1pt} 
\renewcommand{\footrulewidth}{1pt}
\setlength{\arrayrulewidth}{1pt}
\setlength{\columnsep}{6.5mm}

\newcommand{\lambdabar}{\lambda\kern-.5em\raise.5ex\hbox{--}}
\newcommand{\Ninej}[9]{
\left\{
\begin{array}{ccc}
 {#1} & {\!\!\!#2} & {\!\!\!#3}\cr
 {#4} & {\!\!\!#5} & {\!\!\!#6}\cr
 {#7} & {\!\!\!#8} & {\!\!\!#9}\cr
\end{array}
\right\}}
\newcommand{\Cleb}[6]{C^{{\,#1}{\,#2}{\,#3}}
  _{{\,#4}{\,#5}{\,#6}} }
\newcommand{\Sixj}[6]{
\left\{
\begin{array}{ccc}
 {#1} & {\!\!\!#2} & {\!\!\!#3}\cr
 {#4} & {\!\!\!#5} & {\!\!\!#6}\cr
\end{array} 
\right\}}
\newcommand{\Threej}[6]{
\left(
\begin{array}{ccc}
 {#1} & {\!\!\!#2} & {\!\!\!#3}\cr
 {#4} & {\!\!\!#5} & {\!\!\!#6}\cr
\end{array} 
\right)}
\newcommand{\bm}{\mathbf}

\twocolumn[
  \begin{@twocolumnfalse}
\noindent\LARGE{\textbf{Photoassociation spectroscopy of ultracold metastable ${}^{3}$He dimers}}
\vspace{0.6cm}

\noindent\large{\textbf{Daniel G. Cocks,$\ddag$\textit{$^{a}$}
Gillian Peach,\textit{$^{b}$} and
Ian B. Whittingham $^{\ast}$\textit{$^{c}$}}}\vspace{0.5cm}

\noindent\large{\textbf{Abstract}}\\
\noindent \normalsize{The bound states of the fermionic
\mbox{${}^{3}$He(2 $^{3}$S$_{1}$)+${}^{3}$He(2 $^{3}$P$_{j}$)}
system, where $j=0,1,2$, are investigated using the recently available  
\textit{ab initio} short-range ${}^{1,3,5}\Sigma^{+}_{g,u}$ and 
${}^{1,3,5}\Pi_{g,u}$ potentials computed by 
Deguilhem \textit{et al.} (\textit{J. Phys. B: At. Mol. Opt. Phys.}, 2009, 
\textbf{42}, 015102).  Single-channel and multichannel calculations have been undertaken
in order to investigate the effects of Coriolis and non-adiabatic couplings.
The possible experimental observability of the theoretical levels is assessed using 
criteria based upon the short-range character of each level and their coupling to 
metastable ground states. Purely long-range levels have been identified and 30 
short-range levels near five asymptotes are suggested for experimental investigation.} 
\vspace{0.5cm}
 \end{@twocolumnfalse}
  ]

\footnotetext{\textit{$^{a}$~School of Engineering and Physical Sciences, James 
Cook University, Townsville 4811, Australia. E-mail: cocks@itp.uni-frankfurt.de}}
\footnotetext{\ddag~Present address: Institut f\"{u}r Theoretische Physik, Johann 
Wolfgang Goethe-Universit\"{a}t, 60438 Frankfurt/Main, Germany}
\footnotetext{\textit{$^{b}$~Department of Physics and Astronomy, University College London, 
London WC1E 6BT, UK. E-mail: g.peach@ucl.ac.uk}}
\footnotetext{\textit{$^{c}$~School of Engineering and Physical Sciences, James 
Cook University, Townsville 4811, Australia. E-mail: ian.whittingham@jcu.edu.au}}

\section{Introduction}

Photoassociation (PA) of ultracold atoms, in which two interacting ultracold 
atoms are resonantly excited by a laser to bound states of the associated 
molecule, is a widely used technique to study the dynamics of ultracold 
collisions in dilute quantum gases. Of particular interest is PA in metastable 
rare gases where novel experimental strategies based upon their large
internal energy can be implemented.

Photoassociation of ultracold bosonic metastable ${}^{4}$He$^{*}$ atoms,
${}^{4}$He(1s 2s ${}^{3}$S), to excited rovibrational bound states that 
dissociate to the ${}^{4}$He(1s 2s ${}^{3}$S) + ${}^{4}$He(1s 2p ${}^{3}$P$_{j}$)
limits, where $j=0,1,2$, has been observed by many groups. 
The observations include over 40 states 
lying within 14 GHz of the $j=2$ asymptote \cite{Hersch00,Kim04,Rijn04}, six 
states within 0.6 GHz of the $j=1$ asymptote \cite{Rijn04} and some purely 
long-range bound states within 1.43 GHz of the $j=0$ asymptote \cite{Leonard03}.
Theoretical analysis of the $j=0$ long-range states using 
single-channel \cite{Leonard04} and multichannel \cite{Venturi03} calculations
based upon long-range Born-Oppenheimer potentials constructed from retarded 
resonance dipole and dispersion interactions gave excellent agreement with the 
measured binding energies. Analysis of the other states had to await the 
availability of short-range \textit{ab initio} ${}^{1,3,5}\Sigma^{+}_{g,u}$ and 
${}^{1,3,5}\Pi_{g,u}$ molecular potentials \cite{DGL05,DLGD09} and was initially
restricted to single-channel calculations \cite{DGL05,DLGD09} which neglect
non-adiabatic and Coriolis couplings. Very recently a detailed theoretical 
analysis of the entire ${}^{4}$He(1s 2s ${}^{3}$S) + ${}^{4}$He(1s 2p ${}^{3}$P$_{j}$)
system has been completed \cite{CWP10}. The role of these couplings was 
investigated using single-channel and multichannel calculations with the input
potentials constructed from the short-range \textit{ab initio} potentials of 
Deguilhem \textit{et al.}\cite{DLGD09} matched onto long-range retarded resonance dipole and dispersion 
potentials. The multichannel calculations also permitted criteria to be established 
for the assignment of the theoretical levels to experimental observations based 
upon the short-range spin character of each level and their couplings to the 
metastable ground states. Excellent agreement was obtained for the numbers of 
observed levels and their binding energies after application of a 1\% increase 
in the slope of the ${}^{5}\Sigma^{+}_{g,u}$ and ${}^{5}\Pi_{g,u}$ potentials near
their inner classical turning point.  

In contrast, PA of fermionic metastable ${}^{3}$He$^{*}$ atoms,
${}^{3}$He(1s 2s ${}^{3}$S), is relatively unexplored although they have been cooled 
and trapped \cite{SMHV04} with comparable densities and temperatures to those of 
${}^{4}$He$^{*}$ atoms. The non-zero $i=1/2$ nuclear spin of ${}^{3}$He$^{*}$ 
gives rise to hyperfine structure with splittings comparable to the fine structure 
splittings of ${}^{4}$He$^{*}$ which has no nuclear spin. Consequently the 
patterns of energy levels is expected to be quite different for the fermionic and 
bosonic systems. A small number of long-range states in ${}^{3}$He$^{*}$ has been 
predicted by Dickinson \cite{Dickinson06} but this was a single-channel calculation,
thereby neglecting Coriolis and non-adiabatic couplings, using only long-range 
van der Waals and retarded resonance dipole interactions. The availability of the 
short-range potentials of Deguilhem \textit{et al.}\cite{DLGD09} now permits a 
detailed theoretical investigation of the fermionic 
${}^{3}$He(1s 2s ${}^{3}$S) + ${}^{3}$He(1s 2p ${}^{3}$P$_{j}$) system similar 
to that undertaken by Cocks \textit{et al.}\cite{CWP10} for the bosonic ${}^{4}$He$^{*}$ system.

In the absence of any observations of bound states in this excited ${}^{3}$He$^{*}$ 
system, we present predictions as to which of our calculated bound states may be
experimentally observable. We assume any experiment will use magnetic trapping
of the ${}^{3}$He$^{*}$ atoms, requiring all atoms to be in the fully stretched 
low-field seeking $f=3/2, m_{f}=3/2$ magnetic substate of the metastable 
2s ${}^{3}$S$_{1}$ level in order to strongly suppress loss through Penning 
ionization. Consequently we assess the experimental observability of each 
excited level in terms of its coupling to this state. In addition, we consider the 
likelihood of ionization losses from these excited levels due to inelastic collisions
in the short-range region.

Atomic units are used, with 
lengths in Bohr radii $a_{0} = 0.0529177209$ nm and energies in Hartree 
$E_{h}=\alpha ^{2}m_{e}c^{2}=27.211384$ eV.

\section{Theory}
\subsection{Multichannel equations}

The formalism for the excited ${}^{3}$He$^{*}$ system requires  
modification of that presented in Cocks \textit{et al.}\cite{CWP10} for the excited ${}^{4}$He$^{*}$ 
system in order to include hyperfine structure. 

The total Hamiltonian for a system of two interacting atoms $i=1,2$ with reduced 
mass $\mu $, interatomic separation $R$ and relative angular momentum $\hat{\bm{l}}$,
which possess both fine structure and hyperfine structure is 
\begin{equation}
\label{cpw1}
\hat{H} = \hat{T} + \hat{H}_\mathrm{rot} + \hat{H}_\mathrm{el} +
\hat{H}_\mathrm{fs} + \hat{H}_\mathrm{hfs}
\end{equation}
where $\hat{T}$ is the kinetic energy operator
\begin{equation}
\label{cpw2}
   \hat{T}=-\frac{\hbar ^{2}}{2\mu R^{2}}\frac{\partial}{\partial R}
   \left( R^{2}\frac{\partial}{\partial R}\right)\,
\end{equation}
and $\hat{H}_{\mathrm{rot}}$ the rotational operator
\begin{equation}
\label{cpw3}
   \hat{H}_{\mathrm{rot}} = \frac{\hat{l}^{2}}{2 \mu R^{2}}. 
\end{equation}
The total electronic Hamiltonian is 
\begin{equation}
\label{cpw4}
\hat{H}_{\mathrm{el}}=\hat{H}_{1}+\hat{H}_{2}+\hat{H}_{12},
\end{equation}
where the unperturbed atoms have Hamiltonians $\hat{H}_{i}$ and their 
electrostatic interaction  is specified by $\hat{H}_{12}$. The terms 
$\hat{H}_{\mathrm{fs}}$ and $\hat{H}_\mathrm{hfs}$  in equation (\ref{cpw1}) 
describe the fine structure and hyperfine structure respectively of the atoms.  

The multichannel equations describing the interacting atoms are obtained from 
the eigenvalue equation
\begin{equation}
\label{cpw5}
\hat{H} |\Psi\rangle = E |\Psi \rangle
\end{equation}
for the total system by expanding the eigenvector in terms of an appropriate basis 
$|\Phi_a\rangle = |\Phi_a (R,q) \rangle$ where $a$ denotes the set of approximate 
quantum numbers describing the electronic-rotational states of the molecule and $q$ denotes the 
interatomic polar coordinates $(\theta ,\varphi)$ and electronic coordinates 
$(\bm{r}_{1},\bm{r}_{2})$. Using the expansion
\begin{equation}
\label{cpw6}
|\Psi \rangle = \sum_a \frac{1}{R} G_a(R) |\Phi_a\rangle
\end{equation}
and forming the scalar product $\langle \Phi_{a^{\prime}}|\hat{H}|\Psi \rangle $
yields the multichannel equations
\begin{equation}
\label{cpw7}
\sum_a \left\{ T^{G}_{a^\prime a}(R) + \left[V_{a^\prime a}(R) - E 
\delta_{a^{\prime}a}\right] G_{a}(R)\right\} = 0\,,
\end{equation}
where
\begin{equation}
\label{cpw8}
T^{G}_{a^{\prime} a}(R) = -\frac{\hbar^{2}}{2 \mu }\langle \Phi_{a^{\prime}}| 
\frac{\partial ^{2}}{\partial R^{2}} G_{a}(R)|\Phi_{a} \rangle
\end{equation}
and
\begin{equation}
\label{cpw9}
V_{a^{\prime} a}(R) = \langle \Phi _{a^{\prime}}| \left[\hat{H}_{\mathrm{rot}} + 
\hat{H}_{\mathrm{el}} + \hat{H}_{\mathrm{fs}} + \hat{H}_{\mathrm{hfs}}\right] 
|\Phi_{a} \rangle .
\end{equation}
We assume the $R$-dependence of the basis states is negligible so that the radial
kinetic energy term is diagonalized:
\begin{equation}
\label{cpw10}
T^{G}_{a^{\prime} a}(R) =  -\frac{\hbar^{2}}{2\mu }
\frac{d^{2}G_{a}}{dR^{2}}\delta_{a a^{\prime}}.
\end{equation}

\subsection{Basis states}

For two colliding atoms with orbital $\hat{\bm{L}}_{i}$, spin $\hat{\bm{S}}_{i}$ 
and nuclear $\hat{\bm{i}}_{i}$ angular momenta, the unsymmetrized body-fixed 
states in the coupling scheme
\begin{equation}
\label{cpw11}
\hat{\bm{j}}_{i}=\hat{\bm{L}}_{i}+\hat{\bm{S}}_{i},  \quad 
\hat{\bm{f}}_{i}=\hat{\bm{j}}_{i}+\hat{\bm{i}}_{i},  \quad
\hat{\bm{f}}= \hat{\bm{f}}_{1}+\hat{\bm{f}}_{2},     \quad
\hat{\bm{T}}=\hat{\bm{f}}+\hat{\bm{l}}
\end{equation}
are (see appendix for details)  
\begin{equation}
\label{cpw12}
|(\gamma_{1}j_{1}i_{1}f_{1})_{A}, (\gamma_{2}j_{2}i_{2}f_{2})_{B}, f, \Omega_{f},  T, m_{T}\rangle  
\end{equation}
where $\gamma_{i}\equiv\{\bar{\gamma}_{i},L_{i},S_{i}\}$, $\bar{\gamma}_{i}$ representing
any other relevant quantum numbers, and $(A,B)$ labels the two nuclei. 
The projections of an angular momentum $\hat{\bm{J}}$ onto the space-fixed 
$Oz$ and inter-molecular axis $OZ$ with orientation $(\theta,\varphi)$ 
relative to the space-fixed frame will be denoted $m_{J}$ and $\Omega_{J}$
respectively.  

In order to construct states symmetrized with respect to the total parity $\hat{P}_{T}$
we note that $\hat{P}_{T}=\hat{P}_{L}\hat{P}_{S}\hat{P}_{i}\hat{X}_{N}$ where 
$\hat{P}_{L}, \hat{P}_{S}, \hat{P}_{i}$ are the inversion operators on the orbital, 
electronic spin and nuclear spin states associated respectively with 
\begin{equation}
\label{cpw12a}
\hat{\bm{L}}=\hat{\bm{L}}_{1} + \hat{\bm{L}}_{2},  \quad
\hat{\bm{S}}=\hat{\bm{S}}_{1} + \hat{\bm{S}}_{2},  \quad
\hat{\bm{i}}=\hat{\bm{i}}_{1} + \hat{\bm{i}}_{2}
\end{equation}
and $\hat{X}_{N}$ permutes
the nuclei labels. The states of total parity are then (see appendix)
\begin{eqnarray}
\label{cpw13}
\lefteqn{|(\alpha_{1})_{A},(\alpha_{2})_{B},f, \phi, T, m_{T};P_{T}\rangle = }
\nonumber  \\
&& N_{P_{T}}\left[|(\alpha_{1})_{A},(\alpha_{2})_{B},f, \phi, T, m_{T}\rangle \right. 
\nonumber  \\ &&
+ \left. P_{T}P_{1}P_{2}(-1)^{f-T}
|(\alpha_{1})_{A},(\alpha_{2})_{B},f, -\phi, T, m_{T}\rangle \right]
\end{eqnarray}
where $\alpha_{i}\equiv\{\gamma_{i},j_{i},i_{i},f_{i}\}$, $P_{i}=(-1)^{L_{i}}$ is the parity
of the atomic state $|L_{i}m_{L_{i}} \rangle $ and $\phi = |\Omega_{f}|=|\Omega_{T}|$.
The normalization constant is $N_{P_{T}}=1/\sqrt{2(1+\delta_{\phi,0})}$. For 
$\phi =0$ equation (\ref{cpw13}) gives the selection rule $P_{T}P_{1}P_{2}(-1)^{f-T}=1$.

The states symmetrized with respect to $\hat{X}_{N}$ are (see appendix)
\begin{eqnarray}
\label{cpw14}
\lefteqn{|\alpha_{1},\alpha_{2},f, \phi, T, m_{T};P_{T}, X_{N} \rangle = }
\nonumber  \\
&& N_{X_{N}}\left[ |(\alpha_{1})_{A},(\alpha_{2})_{B},f, \phi, T, m_{T};P_{T}\rangle \right. 
\nonumber  \\ &&
+ \left. \varepsilon_{N}|(\alpha_{2})_{A},(\alpha_{1})_{B},f, \phi, T, m_{T};P_{T}\rangle \right]
\end{eqnarray}
where $P_{N}=(-1)^{2i_{1}}$ indicates bosonic or fermionic nuclei (where $i_{1}=i_{2}$
is assumed), $N_{i}$ is the number of electrons on atom $i$, the normalization
constant $N_{X_{N}}$ is $1/\sqrt{2(1+\delta_{\alpha_{1},\alpha_{2}})}$ and 
the phase factor is
\begin{equation}
\label{cpw15}
\varepsilon_{N}=P_{N}P_{T}P_{1}P_{2}(-1)^{f_{1}+f_{2}-f+N_{1}N_{2}}.
\end{equation}
For $\alpha_{1}=\alpha_{2}$, equation (\ref{cpw14}) gives the selection rule 
$P_{N}P_{T}P_{1}P_{2}(-1)^{f_{1}+f_{2}-f+N_{1}N_{2}}=1$.

It is convenient to introduce the simplified notation
\begin{eqnarray}
\label{cpw16}
|\alpha_{1},\alpha_{2}, \phi \rangle = N_{X_{N}}
[|(\alpha_{1})_{A},(\alpha_{2})_{B},f, \phi, T, m_{T} \rangle 
\nonumber  \\
+\varepsilon_{N} |(\alpha_{2})_{A},(\alpha_{1})_{B},f, \phi, T, m_{T} \rangle ]
\end{eqnarray}
so that the states (\ref{cpw14}) can then be written 
\begin{eqnarray}
\label{cpw17}
|\alpha_{1},\alpha_{2},f, \phi, T, m_{T};P_{T}, X_{N} \rangle  
=N_{P_{T}}\left[|\alpha_{1},\alpha_{2}, \phi \rangle \right. \nonumber  \\
 + \left. P_{T}P_{1}P_{2}(-1)^{f-T}|\alpha_{2},\alpha_{1}, -\phi \rangle \right] .
\end{eqnarray}

The eigenstates of $\hat{H}_{\mathrm{el}}$ are the body-fixed states arising 
from the couplings $\hat{\bm{L}}=\hat{\bm{L}}_{1}+\hat{\bm{L}}_{2}$,  
$\hat{\bm{S}}=\hat{\bm{S}}_{1}+\hat{\bm{S}}_{2}$ and must be symmetric under the 
action of $\hat{P}_{L}\hat{P}_{S}$:
\begin{eqnarray}
\label{cpw18}
|\gamma_{1}\gamma_{2},L S \Omega_{L} \Omega_{S};w \rangle =
N_{w}\left[|(\gamma_{1})_{A}(\gamma_{2})_{B}, L S \Omega_{L} \Omega_{S} \rangle \right.
\nonumber  \\
+\varepsilon_{w} \left.
|(\gamma_{2})_{A}(\gamma_{1})_{B}, L S \Omega_{L} \Omega_{S} \rangle \right]
\end{eqnarray}
where $N_{w}=1/\sqrt{2(1+\delta_{\gamma_{1},\gamma_{2}})}$, $w=0(1)$ for
\textit{gerade} (\textit{ungerade}) symmetry and
\begin{equation}
\label{cpw18a}
\varepsilon_{w}=(-1)^{w+L_{1}+L_{2}+S_{1}+S_{2}-S+N_{1}N_{2}}P_{1}P_{2}.
\end{equation}
The relationship between the two bases (\ref{cpw16}) and (\ref{cpw18}) is obtained 
using (see appendix)
\begin{eqnarray}
\label{cpw19}
\lefteqn{|\alpha_{1},\alpha_{2}, \phi \rangle = }
\nonumber  \\
& & N_{X_{N}} N_{w} |T, m_{T}, \phi \rangle
\sum_{ij\Omega_{i}\Omega_{j}} \sum_{LS\Omega_{L}\Omega_{S}} 
F^{f_{1}f_{2}f\phi}_{ji\Omega_{j}\Omega_{i}} 
F^{j_{1}j_{2}j\Omega_{j}}_{LS\Omega_{L}\Omega_{S}} 
\nonumber  \\ && \times
\left[(|g \rangle + |u \rangle ) + \varepsilon
(|g \rangle - |u \rangle )\right]
|(i_{1})_{A}(i_{1})_{B}, i \Omega_{i} \rangle 
\end{eqnarray}
where the coupling coefficients $F^{f_{1}f_{2}f\phi}_{ji\Omega_{j}\Omega_{i}}$ and 
$F^{j_{1}j_{2}j\Omega_{j}}_{LS\Omega_{L}\Omega_{S}}$ are given in the appendix 
(the quantum numbers $(L_{i},S_{i},i_{i})$ have been suppressed) and we have
introduced the notation
\begin{equation}
\label{cpw20}
|g \rangle = |\gamma_{1}\gamma_{2}, L S \Omega_{L} \Omega_{S};g \rangle , \quad
|u \rangle = |\gamma_{1}\gamma_{2}, L S \Omega_{L} \Omega_{S};u \rangle 
\end{equation}
for the eigenstates of \textit{gerade} and \textit{ungerade} symmetry. The rotational
states are
\begin{equation}
\label{cpw21}
|T,m_{T}, \phi \rangle = \sqrt{ \frac{2T+1}{4 \pi }} D^{T\,*}_{m_{T},\phi}(\varphi , \theta , 0),
\end{equation}
where $D^{T\,*}_{m_{T},\phi}(\varphi , \theta , 0)$ is the Wigner rotation matrix, and
the phase factor is 
\begin{equation}
\label{cpw22}
\varepsilon = P_{N} P_{T} (-1)^{2i_{1}+2f+i-2S} .
\end{equation}

For the ${}^{3}$He(1s 2s ${}^{3}$S) + ${}^{3}$He(1s 2p ${}^{3}$P$_{j}$) system, 
$\alpha_{1}=(\bar{\gamma}_{1},0,1,1,1/2,f_{1})$ and 
$\alpha_{2}=(\bar{\gamma}_{2},1,1,j_{2},1/2,f_{2})$ and (\ref{cpw19}) reduces to
\begin{eqnarray}
\label{cpw23}
\lefteqn{|\alpha_{1},\alpha_{2}, \phi \rangle  = }
\nonumber  \\
&&|T, m_{T}, \phi \rangle
\sum_{ij\Omega_{i}\Omega_{j}} \sum_{S\Omega_{L}\Omega_{S}} 
(-1)^{1-j_{2}}[ijf_{1}f_{2}Sj_{2}]^{1/2} \nonumber  \\
&& \times \Cleb{j}{i}{f}{\Omega_{j}}{\Omega_{i}}{\phi}
\Cleb{1}{S}{j}{\Omega_{L}}{\Omega_{S}}{\Omega_{j}}
\Ninej{1}{j_{2}}{j}{1/2}{1/2}{i}{f_{1}}{f_{2}}{f}
\Sixj{1}{1}{j_{2}}{1}{j}{S} \nonumber  \\ 
&& \times
\frac{1}{2}\left[(|g \rangle + |u \rangle ) + P_{T}(-1)^{i}(|g \rangle - |u \rangle )\right]
\nonumber  \\ && \times
|(i_{1})_{A}(i_{1})_{B}, i \Omega_{i} \rangle 
\end{eqnarray}
where $\Cleb{j_1}{j_2}{j}{m_1}{m_2}{m}$ is a Clebsch-Gordan coefficient, 
\mbox{\footnotesize{$\Sixj{a}{b}{c}{d}{e}{f}$}} and 
\mbox{\footnotesize{$\Ninej{a}{b}{c}{d}{e}{f}{g}{h}{i}$}} are 
Wigner $6-j$ and $9-j$ symbols respectively and 
$[ab \ldots ]=(2a+1)\times(2b+1)\times \ldots $.

\subsection{Matrix elements}

The multichannel equations (\ref{cpw7}) require the matrix elements of 
$\hat{H}_{\mathrm{rot}}$, $\hat{H}_{\mathrm{el}}$, $\hat{H}_{\mathrm{fs}}$ and
$\hat{H}_{\mathrm{hfs}}$ in the basis (\ref{cpw16}). Using the notation
$|a \rangle = |\Phi_{a}(R,q) \rangle $ where 
$a \equiv\{\alpha_{1},\alpha_{2},f, \phi ,T, m_{T}, P_{T}, X_{N}\}$ then the 
rotation terms are 
\begin{eqnarray}
\label{cpw24}
\langle a^{\prime}|\hat{l^{2}}|a\rangle & = & \hbar^{2} \delta_{\rho , 
\rho^{\prime}}\left\{\left[T(T+1)+f(f+1)-2\phi^{2}\right]
\delta_{\phi^{\prime},\phi} \right. \nonumber \\
&&  \left. 
-\;K^{-}_{Tf\phi}\;\delta_{\phi^{\prime},\phi-1} -
K^{+}_{Tf\phi}\;\delta_{\phi^{\prime},\phi+1} \right\}\,,
\end{eqnarray}
where the Coriolis coupling terms are 
\begin{eqnarray}
\label{cwp25}
K^{\pm}_{Tf\phi} & = & \left[T(T+1) - \phi(\phi \pm 1)\right]^{\frac{1}{2}}
\nonumber \\ && \times
\left[ f(f+1) - \phi (\phi \pm 1)\right]^{\frac{1}{2}}
\end{eqnarray}
and $\rho$ denotes the set of quantum numbers
$\{\alpha_{1},\alpha_{2},f,T, m_{T}, P_{T}\}$.

The electronic matrix elements can be expressed in terms of the Born-Oppenheimer
(BO) molecular potentials ${}^{2S+1}\Lambda^{\sigma}_{w}(R)$, where 
$\Lambda =|\Omega_{L}|$ and  $\sigma$ is the symmetry of the electronic wave function 
with respect to reflection through a plane containing the internuclear axis, 
using
\begin{equation}
\label{cpw26} 
\hat{H}_{\mathrm{el}}|\gamma_{1}\gamma_{2},L S \Omega_{L} \Omega_{S};w \rangle =
\left[{}^{2S+1}\Lambda^{\sigma}_{w}(R)+ E_{\Lambda S}^{\infty}\right]
|\gamma_{1}\gamma_{2},L S \Omega_{L} \Omega_{S};w \rangle 
\end{equation}
where $E_{\Lambda S}^{\infty}$ is the asymptotic energy of the state. The result is (see appendix)
\begin{eqnarray}
\label{cpw27}
\lefteqn{\langle a^{\prime}|\hat{H}_{\mathrm{el}}|a\rangle  = }
\nonumber  \\ &&
\delta_{\eta , \eta^{\prime}}
\sum_{jj^{\prime}iS}\sum_{\Omega_{L}\Omega_{i}} (-1)^{j_{2}+j_{2}^{\prime}+j+j^{\prime}}
[f_{1}f_{1}^{\prime}f_{2}f_{2}^{\prime}j_{2}j_{2}^{\prime}ff^{\prime}]^{1/2}
\nonumber  \\
&& \times [Sijj^{\prime}]
\Threej{j}{i}{f}{\Omega_{j}}{\Omega_{i}}{-\phi}
\Threej{j^{\prime}}{i}{f^{\prime}}{\Omega_{j}}{\Omega_{i}}{-\phi} 
\nonumber  \\ && \times
\Threej{1}{S}{j}{\Omega_{L}}{\Omega_{S}}{\Omega_{i}-\phi}
\Threej{1}{S}{j^{\prime}}{\Omega_{L}}{\Omega_{S}}{\Omega_{i}-\phi}
\nonumber \\ && \times
\Ninej{1}{j_{2}}{j}{1/2}{1/2}{i}{f_{1}}{f_{2}}{f}
\Ninej{1}{j_{2}}{j}{1/2}{1/2}{i}{f_{1}^{\prime}}{f_{2}^{\prime}}{f}
\nonumber  \\ && \times
\Sixj{1}{1}{j_{2}}{1}{j}{S} \Sixj{1}{1}{j_{2}^{\prime}}{1}{j^{\prime}}{S}
\nonumber \\ && \times
\frac{1}{2}\left\{ {}^{2S+1}\Lambda^{+}_{g}(R)+ {}^{2S+1}\Lambda^{+}_{u}(R)
+2E_{\Lambda S}^{\infty}
\right. \nonumber  \\ && \left. 
+P_{T}(-1)^{i}\left[{}^{2S+1}\Lambda^{+}_{g}(R)- {}^{2S+1}\Lambda^{+}_{u}(R)
\right] \right\}
\end{eqnarray}
where \mbox{\footnotesize{$\Threej{a}{b}{c}{d}{e}{f}$}} is a Wigner $3-j$ coefficient,
$\Omega_{S}=\phi -\Omega_{i}-\Omega_{L}$ and $\eta$ denotes the set of 
quantum numbers $\{\gamma_{1},\gamma_{2},\phi,T,m_{T}, P_{T}\}$. 
This equation differs from that given by Dickinson \cite{Dickinson06} by an 
overall phase factor $(-1)^{1-i-\Omega_{i}}$ and the phase of the 
$\Lambda_{g}-\Lambda_{u}$ term.

The matrix elements of the fine structure and hyperfine structure are best expressed
in the basis 
\begin{equation}
\label{cpw28}
|\alpha_{i}, m_{f_{i}}\rangle = \sum_{m_{j_{i}}m_{i_{i}}}\sum_{m_{L_{i}}m_{S_{i}}} 
\Cleb{j_{i}}{i_{i}}{f_{i}}{m_{j_{i}}}{m_{i_{i}}}{m_{f_{i}}}
\Cleb{L_{i}}{S_{i}}{j_{i}}{m_{L_{i}}}{m_{S_{i}}}{m_{j_{i}}} 
|\gamma_{i},m_{L_{i}},m_{S_{i}}\rangle |i_{i},m_{i_{i}}\rangle .
\end{equation}
For convenience we omit the label $m_{f_{i}}$ from these states as the matrix 
elements of $\hat{H}_{\mathrm{fs}}$ and $\hat{H}_{\mathrm{hfs}}$ are independent 
of $m_{f_{i}}$ due to rotational invariance. We assume that the fine 
structure is independent of $R$ and exclude couplings to the singlet atomic 
state $S_{i}=0$ so that its contribution is
\begin{equation}
\label{cpw29}
\langle \alpha^{\prime}_{i} |\hat{H}_{\mathrm{fs}}|
\alpha_{i} \rangle =
\delta_{\alpha_{i}, \alpha^{\prime}_{i}}\;
\Delta E_{\gamma_{i}j_{i}}^{\mathrm{fs}}.
\end{equation}
The fine structure splitting $\Delta E_{\gamma_{1}j_{1}}^{\mathrm{fs}}$ for the 
2s$\, {}^{3}$S$_{1}$ level vanishes and the splittings $\Delta 
E_{\gamma_{2}j_{2}}^{\mathrm{fs}}$ for the 2p$\,{}^3$P$_0$ and 2p$\,{}^3$P$_1$ 
states relative to the 2p$\,{}^3$P$_2$ level are $31.9088$ GHz and $2.2922$ GHz 
respectively \cite{Wu07}.  

Matrix elements for the hyperfine structure have been obtained by 
Hinds \textit{et al.}\cite{Hinds85} and Wu and Drake \cite{Wu07}.
We choose to use the expression of Wu and Drake but exclude couplings to the $S_{i}=0$
atomic states. The matrix elements are therefore 
\begin{eqnarray}
\label{cpw30}
\lefteqn{\langle \alpha^{\prime}_{i}|\hat{H}_{\mathrm{hfs}}|
\alpha_{i}  \rangle = }
\nonumber  \\
&& \delta_{\gamma_{i},\gamma_{i}^{\prime}}\; W^{i_i f_i}_{j_i j_i^{\prime}}
\left[C_{S_i}\sqrt{6}(-1)^{L_i+j_i^{\prime}}X_{S_i}
\Sixj{S_i^{\prime}}{j_i^{\prime}}{L_i}{j_i}{S_i}{1}\right. 
\nonumber  \\ &&
-D_{S_i}(-1)^{j_i+S_i+M}\Sixj{L_i}{j_i^{\prime}}{S_i}{j_i}{L_i}{1}
\Threej{L_i}{1}{L_i}{-M}{0}{M}^{-1} 
\nonumber  \\ &&
+ \left. E_{S_i}\frac{12}{\sqrt{5}}(-1)^{S_i-L_i+M}X_{S_i}
\right.  \nonumber  \\ && \times  \left.
\Ninej{L_i}{L_i}{2}{S_i}{S_i}{1}{j_i^{\prime}}{j_i}{1} 
\Threej{L_i}{2}{L_i}{-M}{0}{M}^{-1} \right]
\end{eqnarray}
where these expressions are to be evaluated with $M=L_i$,
\begin{equation}
\label{cpw31}
W^{i_i f_i}_{j_i j_i^{\prime}}=(-1)^{j_i+i_i+f_i}i_{i} [j_{i}j_i^{\prime}]^{1/2}
\Sixj{f_i}{i_i}{j_i^{\prime}}{1}{j_i}{i_i}
\Threej{i_{i}}{1}{i_i}{-i_i}{0}{i_i}^{-1}
\end{equation}
and
\begin{equation}
\label{cpw32}
X_{S_i}=-(2S_i+1)\Sixj{1/2}{S_i}{1/2}{S_i}{1/2}{1} .
\end{equation}
The hyperfine structure parameters (in MHz) are \cite{Wu07}
\begin{equation}
\label{cpw33}
C_{1}= -4283.85, \;\; D_{1} = -28.145,
\; \; E_{1}= 7.126.
\end{equation}
The inclusion of hyperfine structure using (\ref{cpw30}) couples states with the same 
$L_{i}$, $S_{i}$ and $f_{i}$ but different $j_{i}$ and for the He 2p$\,{}^{3}$P 
manifold the states $(j,f)=(0,1/2)$ and (1,1/2) are significantly coupled as are
the pair $(j,f)=(1,3/2)$ and (2,3/2). The eigenvalues of 
$\hat{H}_\mathrm{fs} + \hat{H}_\mathrm{hfs}$ 
give the following energies for the hyperfine levels expressed relative to the
state$j=2,f=5/2$: 0, 1780.851, 6292.906, 6961.065 and 34385.941 MHz. The 
eigenvectors give the mixing coefficients which are then used to modify the 
purely algebraic transformation to the hyperfine case given by equation (\ref{cpw28}).
For the hyperfine structure of the 2s$\,{}^3$S level we adopt the 
splitting of $6739.701177$ MHz as measured by Zhao \textit{et al.}\cite{Zhao91}.
This data then gives the ten asymptotic energies $E_{N}^{\infty}$ of the separated 
pairs of atoms as $0$, $1780.851$, $6292.906$, $6739.701$, $6961.065$, $8520.552$, 
$13032.607$, $13700.766$, $34385.941$ and $41125.642$ MHz. 

We assume that the fine- and hyperfine- structure of the individual atoms is not affected 
by their participation within the dimer, so that we may write
\begin{eqnarray}
\lefteqn{\langle a^\prime | \hat{H}_\mathrm{fs} + \hat{H}_\mathrm{hfs} | a \rangle = }
\nonumber \\
&& \delta_{a^\prime a} ( \Delta E^\mathrm{fs}_{\gamma_1 j_1}+ \Delta E^\mathrm{fs}_{\gamma_2 j_2} ) 
+ \delta_{\sigma^\prime , \sigma} 
(\langle \alpha_1^\prime| \hat{H}_\mathrm{hfs} | \alpha_1\rangle \delta_{\alpha_2^\prime ,\alpha_2} \nonumber \\
&& + \langle \alpha_2^\prime| \hat{H}_\mathrm{hfs} | \alpha_2\rangle \delta_{\alpha_1^\prime ,\alpha_1})
\end{eqnarray}
where $\sigma$ denotes the set of quantum numbers $\{f, \phi, T, m_T, P_T\}$.

The total matrix element $V_{a^{\prime}a}(R)$ is therefore diagonal in $\{T,P_{T}\}$
and independent of $m_{T}$. The $m_{T}$-degenerate discrete multichannel 
eigenenergies of (\ref{cpw7}) are then $E_{T,P_{T};v}$ where $v$ 
labels the rovibrational levels.

\subsection{Single-channel approximation}

The single-channel approximation involves the neglect of the Coriolis couplings in 
(\ref{cpw24}) and non-adiabatic couplings in the kinetic energy term. At each value 
of $R$ the single-channel potential is formed by diagonalizing the matrix:
\begin{equation}
\label{cpw34}
V_{a\prime a}^{\phi} = \langle a^{\prime} | \hat{H}_{\mathrm{el}}|a 
\rangle + \langle a^{\prime}|(\hat{H}_{\mathrm{fs}}
+\hat{H}_{\mathrm{hfs}})|a \rangle
+\frac{\langle a^{\prime} |\hat{l}^{2} |a \rangle_{\phi}} {2 \mu R^{2}} \,,
\end{equation}
where $\langle a^{\prime} |\hat{l}^{2} |a \rangle _{\phi}$ is the part of 
(\ref{cpw24}) diagonal in $\phi$.  The corresponding $R$-dependent 
eigenvectors are
\begin{equation}
\label{cpw35}
|n \rangle = \sum_{a} C_{an} (R) |a \rangle
\end{equation}
and the adiabatic potential is given by $V^\mathrm{adi}_n(R) = 
\sum_{a^{\prime}a}C^{-1}_{a^{\prime}n}V^{\phi}_{a^{\prime}a}C_{an}$.
Each channel
$|n \rangle $ can be labelled with the notation $\{\phi, T,m_{T},P_{T}\}$.
 
The adiabatic eigenvalue equation for the rovibrational eigenstates 
$|\psi_{n,v} \rangle = R^{-1}G_{n,v}(R)|i\rangle$, 
where $n=\{\phi , T, m_{T}, P_{T} \}$, is then obtained by neglecting the 
off-diagonal (non-adiabatic) couplings between different single-channel states
in the kinetic energy term so that
\begin{equation}
\label{cpw36}
\langle n^{\prime}|\hat{T}\frac{1}{R}G_{n,v}(R)|n \rangle = 
-\frac{\hbar^{2}}{2\mu R} \frac{d^{2}G_{n,v}}{dR^{2}}\delta_{n, n^{\prime}}.
\end{equation}
The radial eigenvalue equation for the rovibrational states is then
\begin{equation}
\label{cpw37}
\left[ -\frac{\hbar^2}{2\mu }\frac{d^2}{d R^2} + V^\mathrm{adi}_{n}(R) 
- E_{n,v} \right] G_{n,v}(R) = 0\,.
\end{equation}

\subsection{Input potentials}

The required Born-Oppenheimer potentials ${}^{1,3,5}\Sigma^{+}_{g,u}$ and 
${}^{1,3,5}\Pi_{g,u}$ were constructed as in Cocks \textit{et al.}\cite{CWP10} 
by matching the \textit{ab initio} short-range potentials of 
Deguilhem \textit{et al.}\cite{DLGD09} onto the long-range 
dipole-dipole plus dispersion potentials 
\begin{eqnarray}
\label{cpw38}
V_{\Lambda }^{\mathrm{long}}(R)= -f_{3\Lambda}(R/\lambdabar)
C_{3\Lambda}/R^{3} - C_{6\Lambda}/R^{6}
\nonumber  \\ - C_{8\Lambda}^\pm / R^{8} - 
C_{9\Lambda} /R^{9} - C_{10\Lambda}/R^{10},
\end{eqnarray}
where $f_{3\Lambda}$ is an $R$- and $\Lambda$-dependent retardation 
correction~\cite{Meath68}, $\lambdabar=\lambda/(2\pi) = 3258.12 a_{0}$ 
where $\lambda$ is the wavelength for the 2s$\,{}^3$S--2p$\,{}^3$P transition and
the parameters $C_{n \Lambda}$ were taken from  Zhang \textit{et al.}\cite{Zhang06}.

Motivated by our study of the ${}^{4}$He$^{*}$ system \cite{CWP10}, we choose to 
vary the quintet potentials through a modification of the slope of the potential 
at the inner classical turning point by introducing a multiplicative 
factor $c$ through the smoothing function
\begin{equation}
\label{cpw39}
V^{\prime}(R) = \left\{
\begin{array} {ll}
V(R) (1+2c) & R \leq R_1 \\
V(R) \left[ 1 + c(1 + \cos a (R - R_1)) \right] & R_1 < R \leq R_2 \\
V(R) & R > R_2
\end{array}
\right. ,
\end{equation}
where $R_1=5\,a_0$, $R_2=10\;a_0$ and $a = \pi / (R_2-R_1)$.  The value 
$c=0.005$ represents a 1\% variation which is quickly turned on through the 
region $5 < R < 10\,a_{0}$. Its effect is to deepen the minimum of the 
attractive ${}^5 \Pi_g$ potential at $R = 5.387\;a_0$ by $0.985\%$ and move 
it to a smaller interatomic separation by $0.003\;a_0$. The depth of the minimum
in the ${}^5 \Sigma^+_u$ potential at $R = 6.268\;a_0$ is increased by $0.851\%$ 
and is moved towards a smaller separation by $0.010\;a_0$.  The other quintet 
potentials ${}^5 \Sigma^+_g$ and ${}^5 \Pi_u$ are not significantly affected 
as they are repulsive.

\section{Results}

\subsection{Calculations}

The coupled-channel equations (\ref{cpw7}) and the single-channel equation 
(\ref{cpw37}) are of the form
\begin{equation}
\label{cpw40}
\left[ \mathbf{I} \frac{d^{2}}{d R^{2}} + \mathbf{Q}(R)\right] 
\mathbf{G}(R) = 0\,,
\end{equation}
where for the case of coupled-channels, $\mathbf{G}$ is the matrix of solutions 
with the second subscript labelling the linearly independent solutions.  
These equations were solved using the renormalized Numerov 
method \cite{Johnson78} with the eigenvalues of the purely bound states determined 
by counting the nodes of the determinant $|\mathbf{G}(R)|$ and the energies of 
resonances lying within open channels by using a search procedure based on 
Cauchy's argument principle applied to the determinant 
$D(E)= |\mathbf{R}_{m}-\hat{\mathbf{R}}^{-1}_{m+1}|$ where $\mathbf{R}_{m}$ and
$\hat{\mathbf{R}}_{m+1}$ are ratio matrices for the outward and 
inward integrations respectively of the renormalized Numerov method. 
Further numerical details are given in Cocks \textit{et al.}\cite{CWP10}. 

\subsection{Observability Criteria}

In order to predict the likelihood that calculated bound levels may appear in 
future experiments, several properties are determined for each bound level or 
resonance that we isolate.  
The simplest of these is the proportion $P_\mathrm{short}$ of wave function 
present at close interatomic distances, defined as $R<20$~$a_0$ and henceforth referred to 
as the short-range region. This property is extremely useful in classifying 
results since ionization losses, which arise from the inelastic collisions
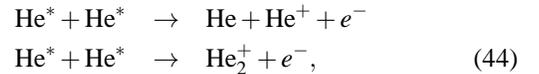
\begin{eqnarray}
\mathrm{He}^* + \mathrm{He}^* & \rightarrow & \mathrm{He} + \mathrm{He}^+ + e^- 
\nonumber \\
\mathrm{He}^* + \mathrm{He}^* & \rightarrow & \mathrm{He}_2^+ + e^- ,
\end{eqnarray}
only occur in the short-range region. As has been observed in bosonic metastable 
helium, there exist indications of purely long-range states in the fermionic dimers investigated 
here, and we define these by $P_\mathrm{short} < 10^{-10}$.

If the level extends into the short-range region then an indication of its propensity 
for ionization is obtained from the proportion $P_\mathrm{str}$ of wave function 
that is in the spin-stretched $S=2$, $i=1$ configuration:
\begin{equation}
\label{cpw40a}
P_\mathrm{str} = \frac{\sum_{a,b} \delta_{S,2} \delta_{i,1} P_{ab}}
{\sum_{a,b} P_{ab}}
\end{equation}
where
\begin{equation}
\label{cpw40b}
P_{ab} = \langle a |b\rangle \int_0^{20\,a_0}\! G_a(R)\;dR
\end{equation}
and $|b\rangle \equiv |\gamma_1\gamma_2,LS\Omega_L \Omega_S w\rangle |(i_1)_A (i_2)_B,i\Omega_i\rangle$
is the complete LS basis state.
The transformation between the bases used here can be found from equation (\ref{cpw19}).
As in the ${}^{4}$He$^{*}$ case, the 
the ionization rate of the dimers is significantly reduced in the spin-stretched 
state \cite{Mcnamara06}. Hence, a large proportion of wave function in the spin-stretched 
state is essential for the level to have a lifetime long enough to be observed 
in experiment.

Finally, for a resonance to be observed in PA experiments, it must be  
strongly coupled by a laser pulse to the metastable manifold 
\mbox{${}^{3}$He(1s 2s $^{3}$S$_{1}$)+${}^{3}$He(1s 2s $^{3}$S$_{1}$)}. 
For radiation of circular polarization $\bm{\epsilon}_{\lambda}$ the
coupling between a metastable dimer state and the excited dimer state 
is due to the interaction 
$\hat{H}_{\mathrm{int}} \sim \bm{\epsilon}_{\lambda}\cdot \bm{d} $
where $\hat{\bm{d}}$ is the molecular dipole moment and is given by
\begin{eqnarray}
\lefteqn{\langle e^\prime | \hat{H}_\mathrm{int} | g \rangle =} \nonumber \\
& &-i (-1)^{\lambda}\;\sqrt{\frac{I}{\varepsilon_{0}c}} \sqrt{\frac{2T+1}{2T^\prime +1}}\; 
C^{T1T^\prime}_{m_T \lambda m_T} \; C^{T1T^\prime}_{\phi \beta \phi^\prime}\; 
\frac{N_{X_{N}}}{2}\; d_\mathrm{at} \nonumber \\
& &\times \sum_{iSj^\prime} \sum_{\Omega_i \Omega_S \Omega_j^\prime} (-1)^{1-j_2^\prime} 
[iS] [j^\prime f_1^\prime f_2^\prime j_2^\prime f_1 f_2]^{1/2} \nonumber \\
& & \times C^{j^\prime i f^\prime}_{\Omega_{j^\prime} \Omega_i \Omega_{f^\prime}} 
C^{1Sj^\prime}_{\beta \Omega_S \Omega_{j^\prime}} C^{Sif}_{\Omega_S \Omega_i \phi} 
\Sixj{1}{1}{j_2^\prime}{1}{j^\prime}{S} 
\nonumber  \\
& & \times 
\Ninej{1}{j_2^\prime}{j^\prime}{1/2}{1/2}{i}{f_1^\prime}{f_2^\prime}{f^\prime} 
\Ninej{1}{1}{S}{1/2}{1/2}{i}{f_1}{f_2}{f}
\end{eqnarray}
where $|g\rangle$ and $|e\rangle$ are basis states corresponding to the 
\mbox{${}^{3}$He(1s 2s $^{3}$S$_{1}$)+${}^{3}$He(1s 2s $^{3}$S$_{1}$)} and 
\mbox{${}^{3}$He(1s 2s $^{3}$S$_{1}$)+${}^{3}$He(1s 2s $^{3}$P$_{j}$)} manifolds 
respectively,  $\beta = \phi^\prime - \phi$ and only those matrix elements with 
$P_T^\prime P_T = -1$ and $|\beta| \leq 1$ are non-zero. The atomic dipole moment 
is given by $d_\mathrm{at}$. Note that the 
metastable spin-stretched state has $P_{T}=+1$ symmetry and can therefore only 
be coupled to excited dimers of $P_{T}=-1$ symmetry.

Finally, we define the quantities
\begin{equation}
\mathcal{A}_\mathrm{str} = \frac{1}{N_{g^\prime}} \sum_{g^\prime a} \langle a| 
\hat{H}_\mathrm{int} |g^{\prime} \rangle \int G_{g^\prime} (R) G_{a,v}(R)\; dR ,
\end{equation}
where $g^\prime$ enumerates all of the $N_{g^\prime}$ spin-stretched metastable 
dimer states, and
\begin{equation}
\mathcal{A}_\mathrm{full} = \frac{1}{N_g} \sum_{g a} \langle a| 
\hat{H}_\mathrm{int} |g \rangle \int G_{g} (R) G_{a,v}(R)\; dR
\end{equation}
where $g$ enumerates all of the $N_{g}$ metastable dimer states. 
The true metastable radial wave functions $G_{g}(R)$ depend upon temperature, 
but in order to extract a single parameter for the observability criteria, we take 
$G_{g}(R) = 1$ as was done in the ${}^4$He$^{*}$ case. 
This is valid up to a constant factor when the metastable wave
functions do not change significantly with temperature.
Although we focus on predicting resonances observable from experiments prepared with 
spin-stretched states in this paper,
due to the overwhelming benefits from reduced trap loss, 
whenever it is convenient we also include 
the likelihood for couplings from other metastable states.
Spin-stretched experiments are best described by the criterion $\mathcal{A}_\mathrm{str}$, 
whereas experiments that do not polarize the metastable gas are best described by 
the criterion $\mathcal{A}_\mathrm{full}$.

\subsection{Single-channel}

The binding energies of long-range states obtained using a single-channel 
calculation are listed in Tables \ref{tbl:01states} and \ref{tbl:23states}. 
The single-channel levels are labelled in terms of $\{T,\phi^{P_{T}}\}$.

Levels which are strongly coupled to the spin-stretched metastable dimer states 
are indicated by a superscript 1.
In the absence of existing experimental data, we use
the same criterion to that obtained for the ${}^{4}$He$^{*}$ case, that is 
$\mathcal{A}_\mathrm{str} > 0.9$~$E_h$.
As these are long-range levels, there is no possibility of ionization and we can 
ignore the $P_\mathrm{str}$ condition. Furthermore, levels that are strongly 
coupled to the unpolarized metastable dimer states are indicated by a 
superscript 2, where the criterion is $\mathcal{A}_\mathrm{full} > 0.9$~$E_h$.  

Of the 159 long-range levels found, 15 have a strong spin-stretched coupling, and 
69 have a strong unpolarized coupling. In addition, there are 151 levels that possess
some short-range character, and also satisfy the observability criteria. Some of these 
are very strongly coupled to the spin-stretched metastable state. However, 
we do not observe these levels once non-adiabatic and Coriolis couplings are turned 
on and so conclude that these levels are unlikely to be observed in experiment.

\begin{table*}
\small
  \caption{\ Single-channel rovibrational binding energies, in units of MHz, of long-range 
$0^{\pm}$ and $1^{\pm}$ states in 
\mbox{${}^{3}$He(2 $^{3}$S$_{1}$)+${}^{3}$He(2 $^{3}$P$_{j}$)}. Energies given are relative
to the energy of the specified asymptote. The superscripts 1 and 2 indicate those
states which satisfy the strong coupling conditions $\mathcal{A}_\mathrm{str} > 0.9$~$E_h$
and $\mathcal{A}_\mathrm{full} > 0.9$~$E_h$ respectively.}
  \label{tbl:01states}
  \begin{tabular*}{\textwidth}{@{\extracolsep{\fill}}lllllllll}
    \hline
	Symmetry & State No. & Asymp. No. & $v/T$ & 0 & 1 & 2  & 3 &  4  \\
    \hline
	$0^{+}$ & 5 & 3 & 0 & 904.113 & & 823.629 & & 639.703 \\
		& &  & 1 & 183.326 & & 127.625 & & 9.53314 \\
		& 6 & 4 & 0 & 347.642 & & 262.537$^2$ & & 75.7138$^2$ \\
		& &  & 1 & 10.8473 & & & & \\
		& 10 & 6 & 0 & & 1422.19 & & 1278.29 & \\
		&  &  & 1 & & 467.747 & & 366.810 & \\
		& &   & 2 & & 62.6480 & & 11.3740 & \\
		& 11 & 8 & 16 & & 52.8477$^2$ & & 27.0296 & \\
		& &   & 17 & & 11.6235 & & &  \\
		& 12 & 9 & 0 & & 202.645$^2$ & & 52.0862$^2$ & \\
		&  &  & 1 & & 13.4273 & & & \\
	$0^{-}$ & 7 & 5 & 0 & 6.10337 & & & & \\
		& 10 & 7 & 0 & 374.065   & 1425.59 & 296.578$^2$ & 1271.97$^2$ & 126.120$^2$  \\
		&  &  & 1 & 40.4103   & 579.160 & 8.45110 & 449.478 &   \\
		& &   & 2 &           & 173.330 &         & 99.6096 &  \\
		&  &  & 3 &           & 37.9230 & & & \\
		& 11 & 8 & 16 &           & 940.315$^2$ &         & 815.427$^2$ &  \\
		&  &  & 17 &           & 503.466$^2$ &         & 417.433 &  \\
		&  &  & 18 &           & 227.478$^2$ &         & 172.257 &  \\
		&  &  & 19 &           & 86.2920$^2$ &         & 55.4097 &  \\
		&  &  & 20 &           & 24.2101$^2$ &         & 9.30074 &  \\
		& 12 & 9 & 16 &           & 500.709$^2$ &         & 319.184$^2$ &  \\
		&  &  & 17 &           & 131.414 &         & 47.6889 &  \\
		&  &  & 18 &           & 19.7607 &         &         &  \\
		& 13 & 10 & 0 &           & 741.860$^2$ &         & 547.258$^2$ &  \\
		&  &  & 1 &           & 233.251 &         & 130.685$^2$ &  \\
		&  &  & 2 &           & 51.1549 &         & 11.7006 &  \\
		&  &  & 3 &           & 5.72522 &         &         &  \\
   $1^{+}$  & 17 & 7 & 0 &           & 1269.28 & 1214.50$^2$ & 1132.90 & 1025.19  \\
		&  &  & 1 &           & 405.296 & 366.915 & 310.695$^2$ & 238.337  \\
		&  &  & 2 &           & 74.8580 & 55.8199 & 29.9220 & 1.87478  \\
		& &   & 3 &           & 6.48198 &         &         &   \\
		& 18 & 8 & 0 &           & 918.791$^2$ & 869.335$^2$ & 796.261$^2$ & 701.007 \\
		&  &  & 1 &           & 432.677$^2$ & 401.217 & 355.166$^2$ & 295.970 \\
		&  &  & 2 &           & 180.375$^2$ & 160.938 & 133.043$^2$ & 98.3456  \\
		&  &  & 3 &           & 62.5650$^2$ & 52.0562 & 37.1800$^2$ & 20.8902  \\
		&  &  & 4 &           & 15.6929 & 10.8815 & & \\
		& 20 & 9 & 0 &           & 438.234$^2$ & 366.110$^2$ & 264.239$^2$ & 141.524$^2$  \\
		&  &  & 1 &           & 106.356 & 73.1955$^2$ & 31.0864 & \\
		&  &  & 2 &           & 14.2935 & 5.04058 & & \\
		& 22 & 10 & 5 &           & 91.4710$^2$ & 58.1199$^2$ & 16.2073$^2$ &  \\
		&  &  & 6 &           & 8.96832 & 1.27172 & & \\
   $1^{-}$  & 11 & 4 & 0 &           & 526.589$^1$ & 465.051$^{1,2}$ & 374.249$^{1,2}$ & 256.239 \\
		&  &  & 1 &           & 31.3952 & 6.28362$^{1,2}$ & & \\
		& 12 & 5 & 0 &           & 30.9200$^{1,2}$ & & & \\
                & 17 & 7 & 0 &           & 342.560$^2$ & 290.731$^{1,2}$ & 214.234$^{1,2}$ & 114.816$^2$ \\
     \hline
  \end{tabular*}
\end{table*}

\begin{table*}
\small
  \caption{\ Single-channel rovibrational binding energies, in units of MHz, of long-range 
$2^{\pm}$ and $3^{\pm}$ states in 
\mbox{${}^{3}$He(2 $^{3}$S$_{1}$)+${}^{3}$He(2 $^{3}$P$_{j}$)}.
Energies given are relative to the energy of the specified asymptote. 
The superscripts 1 and 2 indicate those
states which satisfy the strong coupling conditions $\mathcal{A}_\mathrm{str} > 0.9$~$E_h$
and $\mathcal{A}_\mathrm{full} > 0.9$~$E_h$ respectively.}
  \label{tbl:23states}
  \begin{tabular*}{\textwidth}{@{\extracolsep{\fill}}lllllllll}
    \hline
    Symmetry & State No. & Asympt. No. & $v/T$ & 0 & 1 & 2  & 3 &  4  \\
    \hline
   $2^{+}$  & 6 & 3  & 0 &           &         & 1263.59 & 1167.37 & 1041.69 \\
		& &   & 1 &           &         & 483.739 & 400.323$^2$ & 292.599$^2$ \\
		& &   & 2 &           &         & 24.8937 & & \\
		& 7 & 4  & 0 &           &         & 1119.79$^2$ & 1013.79$^2$ & 875.239$^2$ \\
		&  &  & 1 &           &         & 436.459$^2$ & 376.891$^2$ & 301.106$^2$ \\
		&  &  & 2 &           &         & 143.028 & 105.268 & 58.7279$^2$ \\
		&  &  & 3 &           &         & 10.5484 & & \\
		& 8 & 4 & 0 &           &         & 524.832$^2$ & 441.585$^2$ & 332.271$^2$ \\
   $2^{-}$  & 12 & 8 & 0 &           &         & 932.947$^2$ & 853.811$^{1,2}$ & 749.864$^{1,2}$ \\
		&  &  & 1 &           &         & 325.882$^2$ & 278.283$^2$ & 217.474$^{1,2}$ \\
		&  &  & 2 &           &         & 95.8030 & 71.0373 & 41.3769$^{1,2}$ \\
		&  &  & 3 &           &         & 19.9633 & 10.0466 &  \\
                & 13 & 9 & 0 &           &         & 87.8731$^{1,2}$ & 35.2247$^{1,2}$ &  \\
		&  &  & 1 &           &         & 5.25868 & & \\
   $3^{+}$  & 4 & 4 & 0 &           &         &         & 1623.55 & 1470.74  \\
		&  &  & 1 &           &         &         & 568.235 & 468.826  \\
		&  &  & 2 &           &         &         & 137.420 & 81.9036  \\
		&  &  & 3 &           &         &         & 0.85611 &  \\
   $3^{-}$  & 4 & 4 & 16 &           &         &         & 643.992$^{1,2}$ & 551.467$^{1,2}$  \\
		& &   & 17 &           &         &         & 150.116$^1$ & 97.8055  \\
			&	&    & 18 &           &         &         & 1.20481 &   \\   
\hline
  \end{tabular*}
\end{table*}

\subsection{Multichannel} 

With all couplings included in the calculation, only those levels beneath the 
lowest asymptote are true bound states. In contrast to the situation in 
${}^{4}$He$^{*}$, most of the levels lie above the lowest asymptote and, due to 
couplings to open channels, these higher lying levels almost always acquire a 
finite lifetime due to predissociation. 
These resonances possess complex energies, where the imaginary component represents 
the resonance width, and are more difficult to isolate.  As our search routine
based on Cauchy's argument principle requires many solutions of the differential 
equations (\ref{cpw7}), we restrict the predissociation width to be less than 
100~MHz and only search within 2~GHz of the asymptotic energies that result from 
diagonalization of the hyperfine structure.  Additionally, we match only at two 
points, $100$ and $300$~$a_0$, which may exclude a few levels from our search, 
although it can be argued on the basis of spin-conservation of the laser coupling 
that resonances which exist solely inside this distance will very likely ionize 
and hence will not be observed in experiment.

Beneath the lowest asymptote we find bound levels with only very weak coupling 
strengths. We therefore focus on the resonances that were successfully isolated.
As these levels are not purely long-range, we must also consider the effect of ionization
which reduces the level's lifetime and hence observability. In our previous investigation
of ${}^{4}$He$^{*}$ we imposed a criterion of $P_\mathrm{str} > 87.5\%$.
However, although a large number of resonances were found in ${}^{3}$He$^{*}$ using the above method, very 
few satisfy the same observability criteria as ${}^{4}$He$^{*}$. 
In Table \ref{tbl:02resonances} 
we instead list the 30 resonances that are most likely to be observed in experiment, 
grouped by the nearest fine-structure asymptote.

In contrast to the purely long-range levels in the $0_u^+$, $J=1$ potentials of 
${}^{4}$He$^{*}$, we do not find any single-channel long-range bound levels in 
the ${}^{3}$He$^{*}$ potentials that remain bound after the inclusion of 
couplings to all accessible states, nor do we find any multichannel levels that can 
be described purely in terms of single-channel potentials. 
Again we must emphasize that the relative coarseness of the approach here, necessitated by the
large basis sets, may result in some important levels not being detected.
Additionally, for the remaining resonances with short-range character, very few possess 
strong coupling strengths to the metastable manifold. We do note that there are some particular
resonances which stand out in that their short-range spin-stretch character is 
high with $P_\mathrm{str}>80\%$. It is these levels that we believe will be the most 
likely to be observed in experiment. We also note that the majority of resonances appear 
to be dominated both by $T=1$ and by a projection of $\phi = 1$.

\begin{table*}
\small
  \caption{\ Energies, in units of MHz, of resonances in
\mbox{${}^{3}$He(2 $^{3}$S$_{1}$)+${}^{3}$He(2 $^{3}$P$_{j}$)} that are most 
likely to be observable in experiment. Energies given are relative to the 
specified asymptotic energy $E_{N}^{\infty}$. The predissociation width $\Gamma_\mathrm{pre}$, short-range 
spin-stretched character $P_\mathrm{str}$, coupling strength 
$\mathcal{A}_\mathrm{str}$ and largest contributing basis of $\phi$ are listed 
for each level.}
  \label{tbl:02resonances}
  \begin{tabular*}{\textwidth}{@{\extracolsep{\fill}}lllllll}
    \hline
    $T$ & $P_T$ & $E$ (MHz) & $\Gamma_\mathrm{pre}$ (MHz) & $P_\mathrm{str}$ (\%) 
& $\mathcal{A}_\mathrm{str}$ ($E_h$) & $\phi$  \\
    \hline
	\multicolumn{7}{l}{$E_{2}^{\infty}=1780.85$ MHz} \\
	$2$ & $-1$ & $-1283.40$ & $15.32$ & $49.6$ & $0.372$ & 0 \\ 
	$1$ & $-1$ & $-705.27$ & $71.56$ & $82.0$ & $0.177$ & 1 \\
	$1$ & $-1$ & $-301.47$ & $33.13$ & $90.2$ & $0.267$ & 1 \\
    $2$ & $-1$ & $-110.16$ & $19.02$ & $44.5$ & $0.375$ & 0 \\
	$1$ & $-1$ & $-71.68$ & $5.15$ & $64.4$ & $0.280$ & 1 \\

	\multicolumn{7}{l}{$E_{3}^{\infty}=6292.91$ MHz} \\
	$2$ & $-1$ & $-1951.86$ & $69.9$ & $47.3$ & $0.278$ & 0 \\
	$1$ & $-1$ & $-1808.40$ & $60.4$ & $49.1$ & $0.313$ & 1 \\
	$1$ & $-1$ & $-1179.49$ & $59.8$ & $51.8$ & $0.259$ & 1 \\
	$1$ & $-1$ & $-958.79$ & $60.9$ & $76.1$ & $0.234$ & 1 \\
	$1$ & $-1$ & $-848.05$ & $46.9$ & $70.8$ & $0.234$ & 1 \\
	$1$ & $-1$ & $-812.01$ & $44.2$ & $62.2$ & $0.268$ & 1 \\
	$1$ & $-1$ & $-779.20$ & $44.9$ & $76.9$ & $0.320$ & 1 \\
	$1$ & $-1$ & $-601.37$ & $55.7$ & $49.1$ & $0.285$ & 1 \\
	$2$ & $-1$ & $-499.02$ & $86.6$ & $58.0$ & $0.337$ & 0 \\
	$1$ & $-1$ & $-324.32$ & $54.4$ & $68.1$ & $0.281$ & 1 \\
	$1$ & $-1$ & $-313.77$ & $58.0$ & $75.2$ & $0.312$ & 1 \\

	\multicolumn{7}{l}{$E_{4}^{\infty}=6739.70$ MHz} \\
	$2$ & $-1$ & $-193.32$ & $58.1$ & $51.3$ & $0.301$ & 1 \\
	$2$ & $-1$ & $-186.29$ & $36.0$ & 55.1 & 0.372 &  1 \\
	$1$ & $-1$ & $-38.44$ & $64.7$ & 59.7 & 0.290 &  1 \\
	$1$ & $-1$ & $-11.52$ & $76.6$ & 58.0 & 0.269 & 1 \\

	\multicolumn{7}{l}{$E_{6}^{\infty}=8520.55$ MHz} \\
	$1$ & $-1$ & $-1029.50$ & $20.1$ & 76.5 & 0.204 &  1 \\
	$1$ & $-1$ & $-840.44$ & $24.9$ & 83.8 & 0.212 &  1 \\
	$1$ & $-1$ & $-513.78$ & $42.3$ & 84.8 & 0.204 &  1 \\
	$1$ & $-1$ & $-380.48$ & $53.7$ & 84.0 & 0.190 &  1 \\
	$1$ & $-1$ & $-245.71$ & $55.3$ & 73.0 & 0.213 &  1 \\

	\multicolumn{7}{l}{$E_{7}^{\infty}=13032.61$ MHz} \\
	$1$ & $-1$ & $-1996.78$ & $91.1$ & 76.6 & 0.157 & 0 \\
	$1$ & $-1$ & $-680.18$ & $86.2$ & 80.5 & 0.182 &  1 \\
	$1$ & $-1$ & $-552.73$ & $76.3$ & 74.2 & 0.172 &  1 \\
	$1$ & $-1$ & $-508.69$ & $92.0$ & 76.1 & 0.190 &  1 \\

     \hline
  \end{tabular*}
\end{table*}

\section{Conclusions}

The bound states of the fermionic
\mbox{${}^{3}$He(2 $^{3}$S$_{1}$)+${}^{3}$He(2 $^{3}$P$_{j}$)}
system, where $j=0,1,2$, have been investigated using the recently available  
\textit{ab initio} short-range ${}^{1,3,5}\Sigma^{+}_{g,u}$ and 
${}^{1,3,5}\Pi_{g,u}$ potentials computed by Deguilhem \textit{et al.} \cite{DLGD09}.
Single-channel and multichannel calculations have been undertaken
in order to investigate the effects of Coriolis and non-adiabatic couplings.
In contrast to the situation for the ${}^{4}$He$^{*}$ system \cite{CWP10} where the 
effect of these couplings on the large number of bound levels below the lowest 
asymptote ($j=2$) could be studied, most of the levels for the ${}^{3}$He$^{*}$
lie above the lowest asymptote and become resonances due to couplings to open
channels.

The single-channel long-range levels obtained in the present investigation differ 
significantly from those found by Dickinson \cite{Dickinson06}, both in their patterns
and energies. Dickinson reports nine levels for the $0^{+}$ symmetry, 16 for $0^{-}$,
six for $3^{+}$ and four for $3^{-}$ whereas we find 22 levels for $0^{+}$, 35 for 
$0^{-}$, seven for $3^{+}$ and five for $3^{-}$. We also find numerous levels for
the $1^{\pm}$ and $2^{\pm}$ symmetries for which Dickinson could not find any states. 
These differences are not unexpected as our expression (\ref{cpw27}) for the 
matrix elements of $\hat{H}_{\mathrm{el}}$ differs from that of Dickinson by an 
overall phase factor and the phase of the $\Lambda_{g}-\Lambda_{u}$ term. 
By using the expressions given by Dickinson and with some modification of the values for
hyperfine structure, our single-channel calculations were able to reproduce the 
results of Dickinson to within 5\%.

The possible experimental observability of the theoretical levels has been assessed using 
criteria based upon the short-range character of each level and their coupling to 
metastable ground states.  Although the bound states below the lowest asymptote
and most of the large number of resonances above this asymptote do not satisfy our 
observability criteria we are able to identify some 30 resonances which are 
promising candidates to be observed in experiment.
Unfortunately, the levels that were found in the single-channel calculations were not 
able to be linked to any of the predicted multichannel resonances. This is because we 
only have information regarding resonances that have small predissociation rates, 
instead of for the complete set of states. Hence it is very difficult to observe the 
change of behaviour of a single-channel bound level after the non-adiabatic and 
Coriolis terms are included. In contrast, the ${}^4$He$^{*}$ calculation focused on 
multichannel bound levels which allowed a comparsion between the complete set of  
single-channel and multichannel levels. 
For the short-range levels, this lack of connection implies that the 
non-adiabatic and Coriolis couplings modify the character of the levels such that they are no longer
observable in experiment. However, because the resonance search is costly to perform, we cannot make the
same statement for the purely long-range single-channel levels. Hence, we also recommend 
that future experiments also search for the levels that are marked in Tables 
\ref{tbl:01states} and \ref{tbl:23states} as observable.

\appendix

\section*{Appendix: Basis states and matrix elements}

The unsymmetrized body-fixed (molecular) states in the coupling scheme (\ref{cpw11}) are
\begin{eqnarray}
\label{cpwa1}
\lefteqn{|(\gamma_{1}j_{1}i_{1}f_{1})_{A}, (\gamma_{2}j_{2}i_{2}f_{2})_{B}, f, \Omega_{f},  T, m_{T}\rangle  
=}
\nonumber  \\
&& |T,m_{T}, \Omega_{f} \rangle 
\sum_{\Omega_{f_{1}}\Omega_{f_{2}}}\; 
\sum_{\Omega_{j_{1}}\Omega_{j_{2}}} \;
\sum_{\Omega_{i_{1}}\Omega_{i_{2}}} \;
\sum_{\Omega_{L_{1}}\Omega_{L_{2}}}  \;
\sum_{\Omega_{S_{1}}\Omega_{S_{2}}}
\nonumber  \\ && \times
\Cleb{f_{1}}{f_{2}}{f}{\Omega_{f_{1}}}{\Omega_{f_{2}}}{\Omega_{f}}
\Cleb{j_{1}}{i_{1}}{f_{1}}{\Omega_{j_{1}}}{\Omega_{i_{1}}}{\Omega_{f_{1}}}
\Cleb{j_{2}}{i_{2}}{f_{2}}{\Omega_{j_{2}}}{\Omega_{i_{2}}}{\Omega_{f_{2}}}
\Cleb{L_{1}}{S_{1}}{j_{1}}{\Omega_{L_{1}}}{\Omega_{S_{1}}}{\Omega_{j_{1}}}
\nonumber  \\ && \times
\Cleb{L_{2}}{S_{2}}{j_{2}}{\Omega_{L_{2}}}{\Omega_{S_{2}}}{\Omega_{j_{2}}}
|\gamma_{1}\Omega_{L_{1}}\Omega_{S_{1}}\rangle_{A}|i_{1}\Omega_{i_{1}} \rangle_{A}
\nonumber  \\ && \times
|\gamma_{2}\Omega_{L_{2}}\Omega_{S_{2}}\rangle_{B}|i_{2}\Omega_{i_{2}} \rangle_{B}
\end{eqnarray}
where the transformation between the molecular and space-fixed states is, for example,
\begin{equation}
\label{cpwa2}
|j \Omega_{j} \rangle = \sum_{m_{j}} D^{j}_{m_{j}\Omega_{j}}(\varphi, \theta ,0)
|j m_{j} \rangle .
\end{equation}

The states of the dimer system must be constructed to correctly include the 
symmetries present in the system. Importantly, they must be eigenstates of the 
total parity and nuclear permutation. The total parity operator $\hat{P}_T$ is 
equivalent to the action of $\hat{P}_L \hat{P}_S \hat{P}_i \hat{X}_N$ where 
the action of the inversion operators $\hat{P}_{L}, \hat{P}_{S}$ and $\hat{P}_{i}$ on the orbital, 
electronic spin and nuclear spin space-fixed states respectively is
\begin{eqnarray}
\label{cpwa3}
\hat{P}_{L} |L_{i}m_{L_{i}} \rangle _{A} & = & P_{i} |L_{i}m_{L_{i}}\rangle _{B}, \nonumber  \\
\hat{P}_{S} |S_{i}m_{S_{i}} \rangle _{A} & =  & |S_{i}m_{S_{i}}\rangle _{B}, \nonumber  \\
\hat{P}_{i} |i_{i}m_{i_{i}} \rangle _{A} & =  & |i_{i}m_{i_{i}}\rangle _{B}
\end{eqnarray}
where $P_{i}$ is the parity of the atomic state. The nuclear permutation operator 
$\hat{X}_{N}$ reverses the molecular axis which is equivalent to $A \leftrightarrow B $ and 
$(\theta , \varphi )\rightarrow (\pi - \theta , \varphi + \pi)$. Noting that
\begin{eqnarray}
\label{cpwa4}
\hat{X}_{N} D^{j}_{m_{j},\Omega_{j}}(\varphi , \theta , 0) & = &
D^{j}_{m_{j},\Omega_{j}}(\varphi + \pi , \pi - \theta , 0) 
\nonumber  \\  
 & = & (-1)^{j}D^{j}_{m_{j},-\Omega_{j}}(\varphi , \theta , 0)
\end{eqnarray}
then
\begin{equation}
\label{cpwa5}
\hat{P}_{T} |T,m_{T},\Omega_{f} \rangle = (-1)^{T}|T,m_{T}, -\Omega_{f} \rangle
\end{equation}
and
\begin{eqnarray}
\label{cpwa6}
\lefteqn{\hat{P}_{T}|(\alpha_{1})_{A}, (\alpha_{2})_{B}, f, \Omega_{f},  T, m_{T}\rangle  
=} \nonumber  \\
&& P_{1}P_{2}(-1)^{f-T} 
|(\alpha_{1})_{A}, (\alpha_{2})_{B}, f, -\Omega_{f},  T, m_{T}\rangle  
\end{eqnarray}
where we have introduced the notation $\alpha_{i}=\{\gamma_{i},j_{i},i_{i},f_{i}\}$. 
The eigenstates of $\hat{P}_{T}$ are therefore given by (\ref{cpw13}).

Since $\hat{X}_{N}$ is equivalent to $\hat{P}_{T}\hat{P}_{L}\hat{P}_{S}\hat{P}_{i}$
where
\begin{eqnarray}
\label{cpwa7}
\lefteqn{\hat{P}_{L}\hat{P}_{S}\hat{P}_{i}|(\alpha_{1})_{A}, (\alpha_{2})_{B}, f, 
\Omega_{f},  T, m_{T}\rangle = } \nonumber  \\ 
&& (-1)^{f_{1}+f_{2}-f}P_{1}P_{2}  
|(\alpha_{2})_{A}, (\alpha_{1})_{B}, f, \Omega_{f},  T, m_{T}\rangle  
\end{eqnarray}
then the action of $\hat{X}_{N}$ on the states (\ref{cpw13}) is
\begin{eqnarray}
\label{cpwa8}
\lefteqn{\hat{X}_{N}|(\alpha_{1})_{A},(\alpha_{2})_{B},f, \phi, T, m_{T};P_{T}\rangle 
=} \nonumber  \\
&& P_{T}P_{1}P_{2}(-1)^{f_{1}+f_{2}-f+N_{1}N_{2}}  
\nonumber  \\ &&  \times
|(\alpha_{2})_{A},(\alpha_{1})_{B},f, \phi, T, m_{T};P_{T}\rangle 
\end{eqnarray}
so that the eigenstates of $\hat{X}_{N}$ are (\ref{cpw14}).

The relationship (\ref{cpw19}) between the bases (\ref{cpw16}) and (\ref{cpw18}) is 
obtained by first using 
\begin{eqnarray}
\label{cpwa9}
\lefteqn{|\gamma_{1}\Omega_{L_{1}}\Omega_{S_{1}}\rangle _{A}
|\gamma_{2}\Omega_{L_{2}}\Omega_{S_{2}}\rangle _{B} = }
\nonumber  \\ &&
\sum_{LS\Omega_{L}\Omega_{S}}
\Cleb{L_{1}}{L_{2}}{L}{\Omega_{L_{1}}}{\Omega_{L_{2}}}{\Omega_{L}}
\Cleb{S_{1}}{S_{2}}{S}{\Omega_{S_{1}}}{\Omega_{S_{2}}}{\Omega_{S}}
\nonumber  \\ &&  \times
|(\gamma_{1})_{A}(\gamma_{2})_{B},LS \Omega_{L}\Omega_{S} \rangle  ,
\end{eqnarray}
and expressing sums over Clebsch-Gordan coefficients as $9-j$ symbols to give
\begin{eqnarray}
\label{cpwa10}
\lefteqn{|(\alpha_{1})_{A},(\alpha_{2})_{B},f, \phi, T, m_{T};P_{T}\rangle = } \nonumber  \\
&& |T,m_{T}, \phi \rangle 
\sum_{\Omega_{i_{1}}\Omega_{i_{2}}} \sum_{ij\Omega_{i}\Omega_{j}} \sum_{LS\Omega_{L}\Omega_{S}}
\nonumber  \\ && \times
[ijLSj_{1}j_{2}f_{1}f_{2}]^{1/2} 
\Cleb{j}{i}{f}{\Omega_{j}}{\Omega_{i}}{\phi}
\Cleb{i_{1}}{i_{2}}{i}{\Omega_{i_{1}}}{\Omega_{i_{2}}}{\Omega_{i}}
\nonumber  \\ && \times
\Cleb{L}{S}{j}{\Omega_{L}}{\Omega_{S}}{\Omega_{j}}
\Ninej{j_{1}}{j_{2}}{j}{i_{1}}{i_{2}}{i}{f_{1}}{f_{2}}{f}
\Ninej{L_{1}}{L_{2}}{L}{S_{1}}{S_{2}}{S}{j_{1}}{j_{2}}{j} 
\nonumber  \\ && \times
|(\gamma_{1})_{A}(\gamma_{2})_{B},LS \Omega_{L}\Omega_{S} \rangle 
|i_{1}\Omega_{i_{1}}\rangle _{A} |i_{2} \Omega_{i_{2}}\rangle_{B} .
\end{eqnarray}
Introducing the coupling coefficients, for example,
\begin{eqnarray}
\label{cpwa11}
F_{LS\Omega_{L}\Omega_{S}}^{j_{1}j_{2}j\Omega_{j}} & = & 
[(2L+1)(2S+1)(2j_{1}+1)(2j_{2}+1)]^{\frac{1}{2}}
\nonumber \\ && \times
\Cleb{L}{S}{j}{\Omega_L}{\Omega_S}{\Omega_j}
\Ninej{L_1}{L_2}{L}{S_1}{S_2}{S}{j_1}{j_2}{j} \,
\end{eqnarray}
and using, from (\ref{cpw18}),
\begin{equation}
\label{cpwa12}
|(\gamma_{1})_{A}(\gamma_{2})_{B},LS \Omega_{L}\Omega_{S} \rangle 
= N_{w}(|g \rangle + |u \rangle )
\end{equation}
then
\begin{eqnarray}
\label{cpwa13}
\lefteqn{|(\alpha_{1})_{A},(\alpha_{2})_{B},f, \phi, T, m_{T};P_{T}\rangle = } 
\nonumber  \\
&& |T,m_{T}, \phi \rangle 
\sum_{\Omega_{i_{1}}\Omega_{i_{2}}}  \sum_{ij\Omega_{i}\Omega_{j}} \sum_{LS\Omega_{L}\Omega_{S}}
\Cleb{i_{1}}{i_{2}}{i}{\Omega_{i_{1}}}{\Omega_{i_{2}}}{\Omega_{i}}
\nonumber  \\ &&  \times
F^{f_{1}f_{2}f\phi}_{ji\Omega_{j}\Omega_{i}} \;
F^{j_{1}j_{2}j\Omega_{j}}_{LS\Omega_{L}\Omega_{S}} 
(|g \rangle + |u \rangle )|i_{1}\Omega_{i_{1}}\rangle _{A} |i_{2} \Omega_{i_{2}}\rangle_{B} .
\end{eqnarray}

The state with $A \leftrightarrow B$ is obtained by reordering the angular momenta 
subscripted with 1 and 2 in all Clebsch-Gordan and $9-j$ symbols and using
\begin{equation}
\label{cpwa14}
|(\gamma_{2})_{A}(\gamma_{1})_{B},LS \Omega_{L}\Omega_{S} \rangle 
= N_{w}P_{1}P_{2}(-1)^{L_{1}+L_{2}-L+S_{1}+S_{2}-S}(|g \rangle - |u \rangle )
\end{equation}
to give
\begin{eqnarray}
\label{cpwa15}
\lefteqn{|(\alpha_{2})_{A},(\alpha_{1})_{B},f, \phi, T, m_{T};P_{T}\rangle = }
\nonumber  \\
&&  |T,m_{T}, \phi \rangle 
\sum_{\Omega_{i_{1}}\Omega_{i_{2}}} 
 \sum_{ij\Omega_{i}\Omega_{j}} \sum_{LS\Omega_{L}\Omega_{S}}
(-1)^{f_{1}+f_{2}+f+2i}
\nonumber  \\ && \times 
\Cleb{i_{1}}{i_{2}}{i}{\Omega_{i_{1}}}{\Omega_{i_{2}}}{\Omega_{i}}
F^{f_{1}f_{2}f\phi}_{ji\Omega_{j}\Omega_{i}} \;
F^{j_{1}j_{2}j\Omega_{j}}_{LS\Omega_{L}\Omega_{S}} 
\nonumber  \\ && \times
(|g \rangle - |u \rangle ) |i_{2}\Omega_{i_{2}}\rangle _{A} |i_{1} \Omega_{i_{1}}\rangle_{B}. 
\end{eqnarray}
Forming the combination (\ref{cpw15}) then yields (\ref{cpw19}).

The matrix elements of $\hat{H}_{\mathrm{el}}$ are diagonal in $\phi $. Using 
the explicit states (\ref{cpw23}) then
\begin{eqnarray}
\label{cpwa16} 
\lefteqn{\langle \alpha_{1}^{\prime},\alpha_{2}^{\prime}, \phi^{\prime}|\hat{H}_{\mathrm{el}}
|\alpha_{1}, \alpha_{2}, \phi \rangle = } 
\nonumber  \\
&& \langle T^{\prime},m_{T}^{\prime}, \phi^{\prime}|T, m_{T}, \phi \rangle 
\sum_{i^{\prime}j^{\prime}\Omega_{i}^{\prime}\Omega_{j}^{\prime}}
\sum_{ij\Omega_{i}\Omega_{j}}
\sum_{S^{\prime}\Omega_{L}^{\prime}\Omega_{S}^{\prime}}
\sum_{S\Omega_{L}\Omega_{S}}
(-1)^{j_{2}^{\prime}+j_{2}}
\nonumber  \\ && \times
[i^{\prime}j^{\prime}f_{1}^{\prime}f_{2}^{\prime}S^{\prime}j_{2}^{\prime}
ijf_{1}f_{2}Sj_{2}]^{1/2}
\nonumber  \\  && \times
\Cleb{j^{\prime}}{i^{\prime}}{f^{\prime}}{\Omega_{j}^{\prime}}
{\Omega_{i}^{\prime}}{\phi^{\prime}}
\Cleb{j}{i}{f}{\Omega_{j}}{\Omega_{i}}{\phi}
\Cleb{1}{S^{\prime}}{j^{\prime}}{\Omega_{L}^{\prime}}{\Omega_{S}^{\prime}}{\Omega_{j}^{\prime}}
\Cleb{1}{S}{j}{\Omega_{L}}{\Omega_{S}}{\Omega_{j}}
\nonumber  \\  && \times
\Ninej{1}{j_{2}^{\prime}}{j^{\prime}}{1/2}{1/2}{i^{\prime}}{f_{1}^{\prime}}{f_{2}^{\prime}}
{f^{\prime}}
\Ninej{1}{j_{2}}{j}{1/2}{1/2}{i}{f_{1}}{f_{2}}{f}
\nonumber  \\ && \times
\Sixj{1}{1}{j_{2}^{\prime}}{1}{j^{\prime}}{S^{\prime}}
\Sixj{1}{1}{j_{2}}{1}{j}{S}
\nonumber  \\ && \times
\frac{1}{4}\left[(c_{iT}^{g\prime}\langle g^{\prime}| + c_{iT}^{u\prime}\langle u^{\prime})|
\hat{H}_{\mathrm{el}}|(c_{iT}^{g}|g \rangle + c_{iT}^{u} |u \rangle )\right]
\nonumber  \\ && \times
\langle (i_{1}^{\prime})_{A},(i_{1}^{\prime})_{B}, i^{\prime},\Omega_{i}^{\prime}|
(i_{1})_{A}, (i_{1})_{B},i,\Omega_{i} \rangle
\end{eqnarray}
where $c_{iT}^{g}=1+(-1)^{i}P_{T}$ and $c_{iT}^{u}=1-(-1)^{i}P_{T}$. The orthogonality of
states reduces this to
\begin{eqnarray}
\label{cpwa17}
\lefteqn{\langle \alpha_{1}^{\prime},\alpha_{2}^{\prime}, \phi^{\prime}|\hat{H}_{\mathrm{el}}
|\alpha_{1}, \alpha_{2}, \phi \rangle = } 
\nonumber  \\
&& \delta_{\eta, \eta^{\prime}}
\sum_{j^{\prime}ij} \sum_{\Omega_{i}\Omega_{j}}\sum_{S\Omega_{L}\Omega_{S}}
(-1)^{j_{2}^{\prime}+j_{2}}
[j^{\prime}f_{1}^{\prime}f_{2}^{\prime}j_{2}^{\prime}jf_{1}f_{2}j_{2}]^{1/2}[Si]
\nonumber  \\  && \times
\Cleb{j^{\prime}}{i}{f^{\prime}}{\Omega_{j}^{\prime}}{\Omega_{i}}{\phi}
\Cleb{j}{i}{f}{\Omega_{j}}{\Omega_{i}}{\phi}
\Cleb{1}{S}{j^{\prime}}{\Omega_{L}}{\Omega_{S}}{\Omega_{j}^{\prime}}
\Cleb{1}{S}{j}{\Omega_{L}}{\Omega_{S}}{\Omega_{j}}
\nonumber  \\ && \times
\Ninej{1}{j_{2}^{\prime}}{j^{\prime}}{1/2}{1/2}{i}{f_{1}^{\prime}}{f_{2}^{\prime}}
{f^{\prime}}
\Ninej{1}{j_{2}}{j}{1/2}{1/2}{i}{f_{1}}{f_{2}}{f}
\Sixj{1}{1}{j_{2}^{\prime}}{1}{j^{\prime}}{S}
\nonumber  \\ && \times
\Sixj{1}{1}{j_{2}}{1}{j}{S}
\frac{1}{2}\left[c_{iT}^{g}({}^{2S+1}\Lambda_{g}^{+}(R)+E_{\Lambda S}^{\infty})
\right.
\nonumber  \\ &&  
+ \left. c_{iT}^{u}({}^{2S+1}\Lambda_{u}^{+}+E_{\Lambda S}^{\infty}) \right]
\end{eqnarray}
where $\eta = \{\gamma_{1},\gamma_{2},\phi, T, m_{T}, P_{T}\}$. Similarly we
can show 
\begin{eqnarray}
\label{cpwa18}
\lefteqn{\langle \alpha_{1}^{\prime},\alpha_{2}^{\prime}, -\phi^{\prime}|\hat{H}_{\mathrm{el}}
|\alpha_{1}, \alpha_{2},- \phi \rangle = } 
\nonumber  \\
&& (-1)^{f-f^{\prime}}
\langle \alpha_{1}^{\prime},\alpha_{2}^{\prime}, \phi^{\prime}|\hat{H}_{\mathrm{el}}
|\alpha_{1}, \alpha_{2}, \phi \rangle . 
\end{eqnarray}
The phase factor $(-1)^{f-f^{\prime}}$ is cancelled by the factor 
arising from (\ref{cpw17}), thus ensuring 
the two contributions (\ref{cpwa17}) and (\ref{cpwa18}) interfere constructively.
Conversion of the Clebsch-Gordan coefficients into $3-j$ symbols finally yields
(\ref{cpw27}).

\section*{Acknowledgements}

Our initial interest in this problem was stimulated by some very interesting and 
helpful discussions with the late Professor Alan Dickinson of Newcastle University.
Therefore we would like to take this opportunity to dedicate this paper to his 
memory.

\footnotesize{

\end{document}